\documentclass[10pt,letterpaper]{article}
\usepackage{cogsci}

\cogscifinalcopy 

\usepackage{pslatex}
\usepackage{apacite}
\usepackage{float} 
\usepackage{multirow}
\usepackage[inline]{enumitem}
\usepackage{amsmath}
\usepackage{graphicx}
\usepackage{lipsum}
\usepackage{float}

\title{Why do you take that route?}

 \author{\bf\Large Alimire Nabijiang,\textsuperscript{1}
 Supratik Mukhopadhyay,\textsuperscript{1}
 Yimin Zhu,\textsuperscript{2}\\
 {\bf\Large Ravindra Gudishala,\textsuperscript{3}
 Sanaz Saeidi,\textsuperscript{2}
 Qun Liu\textsuperscript{1}}\\
 \textsuperscript{1}{Department of Computer Science and Engineering}\\
 \textsuperscript{2}{Department of Construction Management}\\
 \textsuperscript{3}{Department of Civil and Environmental Engineering and Louisiana Transportation Research Center}\\
 Louisiana State University, Baton Rouge, LA 70803\\
 \{anabij1, supratik, yiminzhu, ssaeid1, rgudis, qliu14 \}@lsu.edu
 }

\begin{document}
\maketitle

\begin{abstract}
The purpose of this paper is to determine whether a particular context 
factor among the variables that a researcher is  interested  in causally affects the
route-choice behavior of drivers. To our knowledge, there is
limited literature that consider the effects of various factors
on route choice based on causal inference.Yet, collecting
data sets that are sensitive to the aforementioned factors are
challenging and the existing approaches usually take into account only 
the general factors motivating drivers route choice behavior. To fill these gaps, we carried out a study using Immersive Virtual Environment (IVE) tools to elicit drivers’ route choice behavioral
data, covering drivers' network familiarity, education
level, financial- concern, etc, apart from conventional
measurement variables. Having context-aware, high-fidelity
properties, IVE data affords the opportunity
to incorporate the impacts of human-related factors into the
route choice causal analysis and advance a more customizable
research tool for investigating causal factors on path selection
in network routing. This causal analysis provides quantitative
evidence to support drivers’ diversion decision. The study
also provides academic suggestion and reference for investing in 
public infrastructure and developing efficient strategies
and policies to mitigate traffic congestion. 


\end{abstract}

\section{Introduction} 

Route choice refers to the choices of roads among a set of possible alternatives made by human drivers while navigating through an urban area. Route Choice Models estimate the route choices of drivers in an urban setting. Most route choice models connect  characteristics of alternate routes to those selected by the drivers. These models  help in  estimating traffic levels on different routes and thus enable development of effective traffic management strategies that can reduce traffic delays and allow maximum utilization of transportation systems. Existing route choice models use revealed preference behavior to model route choice. The use of revealed choice data limits the accuracy of the prediction as it fails to capture subjective context factors of drivers at individual level. Therefore, it is essential to use a data collection methodology that incorporates  the importance of contextual factors in route choice. 

As commuters we all make choices on which route to use when traveling to work, school, or shopping mall. Most of the times we pick a route that is familiar and also minimizes travel time. However, there is plenty of evidence that as commuters we take routes that do not minimize travel time \cite{ben2004route}. In order to try and explain this route choice behavior, transportation engineers have been studying the route choice behavior of drivers for the past three decades to try and explain it. Transportation researchers have adopted econometric based approaches and used two types of data to mathematically model and rationalize the route choice behavior \cite{prato2009route}.

The data used in modeling route choices is collected by using two approaches. The first approach is based on actual observed route choice behavior that is often labeled as Revealed Preference data. The second approach is based on collecting data from hypothetical choice experiments that is often called as State Choice data \cite{ben2010road}. There are times when both types of data are combined to model and explain route choice behavior. However, the combination of econometric approaches and different data collection methods have yielded mixed results in explaining route choice behavior.

Based on the literature reviewed we believe that there is not much research that have tried to apply causal analysis methods to explain route choice behavior. We believe that by applying causal analysis techniques we can identify root causes that influence route choice and will subsequently allow us to enhance Route Choice models that will better forecast traffic levels on transportation networks and also to better comprehend drivers’ response to route guidance and dynamic message signs.

The main objective of this paper is to conduct causal analysis of route choice behavior using data collected from a Stated Choice Experiment in an Immersive Virtual Environment (IVE). We carried out a study using IVE tools to elicit drivers’ route choice behavioral
data, covering drivers network familiarity, education
level, financial- concern, etc, other than conventional
measurement variables. Having context-aware, high-fidelity
properties, IVE data affords the opportunity
to incorporate the impacts of human-related factors into the
route choice causal analysis and advance a more customizable
research tool for investigating causal factors on path selection
in network routing. This causal analysis provides quantitative
evidence to support drivers’ diversion decision. The study
also provides academic suggestion and reference for investing in 
public infrastructure and developing efficient strategies
and policies to mitigate traffic congestion. 

This paper makes the following contribution:
\begin{itemize} 
\item To the best of our knowledge, the paper presents the first causal analysis of route choice behavior of drivers using data collected from a Stated Choice Experiment in an Immersive Virtual Environment (IVE).
\end{itemize}

\section{Related Work}
Transportation engineers have been studying commuter route choice behavior for four decades now. Engineers  developing route choice models theorized that travel time plays a crucial and important role in  the selection of a route.  Route choice behavior theories began to evolve in  the late eighties and early nineties as engineers' understanding of route choice behavior improved by studying data about  empirical route choice behavior. Pursula and Talvite \cite{pursula1993urban} developed a mathematical route model by postulating that drivers do consider other factors apart from travel time in making a route choice. In \cite{khattak1993commuters}, the authors  discovered that commuters prefer to use habitual routes when traveling in familiar areas as opposed to choosing a route that provides them with maximum utility. Other researchers such as Doherty and Miller \cite{doherty2000computerized} investigating route choice found that apart from travel time, factors such as residential location, familiarity with the route, and employment locations are significant in the route choice process. Deep learning techniques \cite{basu2018deep,basu2015deepsat,lecun2015deep,qun,lv2015traffic,song2016deeptransport} can be used to predict traffic congestion and route choice. However, deep learning models, being opaque,  cannot be used to causally explain drivers' route choice. 

In reviewing the existing research it can be gleaned that transportation researchers have employed two different types of empirical data collection in studying route choice behavior. First, collecting route choice data using observed actual choices and second, collecting route choice data in hypothetical experiments.  Researchers have for the majority of cases used utility maximizing theory to explain route choice behavior that is rooted in econometrics \cite{ben1985discrete}. 

\section{Constructing Graphical Causal Models}
In causal inference, we need a way of formally representing our assumptions about causal relationship within data. Graphical models comes in handy for this purpose. There are  a variety of  ways to depict causal relationship using graphical causal models \cite{spiegelhalter1993bayesian,glymour1999computation,neapolitan2004learning,pearl2009causality}. A graphical model provides a clear way to represent and better understand the causal relationships within a data set \cite{pearl2014probabilistic,alma}. A Causal graph is useful in determining the cause-effects from data by identifying confounding and endogenous selection bias. We also can derive a testable implications from  the graph to test our assumptions \cite{elwert2013graphical}. To construct  a graphical model requires  subject-matter understanding \cite{hernan}.

In our study, to model causal assumptions we carried out an iterative procedure following three steps. We identified the related variables and constructed our pilot causal graph via one-to-one discussions with experts in the field of transportation. In the second step, since a casual graph reveals  testable implications, we tested our assumption to some extent using graphical criteria. In the final step, we evaluated the pilot model discussing with experts again. We modified our graph according to the discussion with experts and results obtained by testing the model against data. After proper adjustments, we finalized the causal model  for  further causal inference procedure.

\section{Data Collection}
Route choice can be influenced by factors, such as, road condition and human-related factors (i.e., driving experience, driver's socio-economic characteristics, and driving behavior and attitudes \cite{Deona}). The current route choice models are calibrated using static contextual conditions and are not generally able to account for accessibility to the nearest freeway, traffic incidents, and road closures due to emergency. Collecting dataset including dynamic contextual factors are challenging and the existing approaches usually take into account the general factors motivating drivers’ route choice behavior. In causal inference of route choice, it is preferable to have as much as data related to contextual factors which have potential influence on drivers' route choice decision. This study conducts experimental scenarios in which specific contextual factors are added in the testing design, using Virtual Reality (VR) platform and a driving simulator.The study, in particular, examines individuals’ diversion tendency onto alternate routes that are induced by traffic condition, journey type, and the impact of social influence while driving in the Interstate 10 (I-10) freeway in Baton Rouge, between the Mississippi River Bridge and College Drive exit. Collecting route choice data in hypothetical experiments facilitated our study by providing various factors information for causal analysis.

 \textbf{IVE Experimental Setting} 
 In this study, we used a driving environment that is designed based on the I-10, starting off the Mississippi River bridge all the way to the College Dr. Along the way, five alternate routes were introduced to the participants — Exits A, B, C, D, and E, the latter of which would be College Drive. Ten experimental scenarios were conducted to produce initial data about drivers’ dynamic route choice behavior, given emerging contextual factors. See Table \ref{table1}. 

 \begin{table}[h]
\begin{center} 
\caption{Contextual Factors Description} 
\label{table1} 
\vskip 0.12in
\begin{tabular}{lc} 
\hline
 Contextual Factors   & Scale \\
\hline
\multirow{3}{*}{Traffic Condition} &Normal\\
  &Medium \\
  &Heavy \\
 \hline    
\multirow{2}{*}{Journey Type}   &Urgent \\
                 &Non-Urgent\\
\hline
\multirow{2}{*}{Social Impact}   &Yes\\
                                 &No\\
\hline
\end{tabular} 
\end{center} 
\end{table}
 Forty-one individuals (20 male and 21 females; age: $31.44 \pm 7.97$) volunteered to participate in the experiment. Prior to the experiment, participants were presented with a questionnaire asking the following items: \begin{enumerate*}[label=\arabic*)]\item demographic characteristics (age, gender, race, education, employment status); \item top concerns while they stuck in the traffic congestion. Their choices included hours of extra travel time, speed reduction, monetized value of delay;\item familiarity with the area; \item socio-economic status (having concerns about spending less money on your gas).
 \end{enumerate*}
 After answering the questionnaires, participants were asked to sit on a stationary chair at a desk with a driving wheel which was placed in front of a  flat screen monitor where the driving simulation would run.  Next, they were invited to practice for a few minutes to get acquainted with driving the simulator. After enough practicing with the driving simulator and becoming comfortable with its environment the research team would assign the participant to the scenarios. See Table \ref{table2}. 
 
 \begin{table}[h]
\begin{center} 
\caption{Experimental Scenarios of the Study} 
\label{table2} 
\vskip 0.12in
\begin{tabular}{ccc} 
\hline
 Traffic Condition  &Journey Type  &Social Impact\\
\hline
Normal & Urgent   &No   \\
Medium     & Urgent  &No \\
Heavy          & Urgent   &No   \\
Medium           & Urgent   &Yes     \\
Heavy     &Urgent     &Yes          \\
Normal & Non-Urgent   &No\\
Medium  &Non-Urgent   &No\\
Heavy    &Non-Urgent    &No\\
Medium   &Non-Urgent    &Yes\\
Heavy     &Non-Urgent    &Yes\\
\hline
\end{tabular} 
\end{center} 
\end{table}
 
 The origin and the destination in all the scenarios were same and each scenario took about two minutes to finish. In each scenario different contextual factor(s) were presented and participants were required to choose their preferred route.  
 
 Each participant was exposed to all the driving scenarios including a baseline scenario. The baseline scenario would collect information about participant’s route choice pattern in a normal traffic and non-urgent bound condition. Each scenario contained 1, 2, or 3 contextual factors. The first contextual factor was the traffic flow which was varied over three levels, i.e., normal, medium, and heavy density. The next factor was the purpose of the trip (journey type) which consisted of a work-bound and home-bound trips; on the work-bound trip, participants were told to consider how important was it to meet the time of arrival commitment, while the home-bound posed no rush to reach the destination. The third factor considered is the impact of other drivers’ route choice, exploring the idea of social influence, that is whether the driver would be influenced by watching other drivers take an exit. 
 
 Dynamic route guidance was presented to the participants where a driver is guided on to routes that will minimize travel time for the overall road network. 
 The scenarios were counterbalanced and played out in a random fashion to avoid behavioral biases due to order effect. 
 
 \section{DAGs, D-Separation, Testable Implication}
\textbf{DAGs} Directed Acyclic Graphs (DAGs) can represent probability distributions of the data and can be considered as causal graphical models under three important assumptions \cite{hernan}. First, we assume that direct causal effect exists between pairs of variables connected by directed edges. Second, we  assume that DAGs satisfy the Causal Markov condition \cite{hernan}. The Causal Markov condition states that a variable is independent of every other variable except its effects conditioned on all of its direct causes \cite{hernan,andlenz}. Mathematically, this is expressed as: 
\begin{equation}
f(V)=\overset{n}{\underset{i=1}{\prod}} f(x_i | pa_i))
\end{equation}

\noindent where $f(V)$ denotes joint probability  mass function over the set of nodes $V$.  The variables $pa_i$ denote the values of the direct causes of variables $x_i$, and $i$ takes values from $1$ to $n$. 

The third condition is Faithfulness. By assuming that a causal graph satisfies Causal Markov condition, we assume that any population produced by this causal graph has the conditional independence relations obtained by applying d-separation. \cite{scheines1997introduction}. 

We can test the assumption (the pilot causal model) by applying d-separation \cite{pearl2014probabilistic}. This allows us to verify if the model fits the data. If the conditional independence test based on data violates the d-separation rule, we can modify original model. Fortunately, d-separation rules spot the flaws locally so we can fix the problems without much effort. We don't need to throw away the model and  start the whole process from scratch.

\noindent \textbf{D-SEPARATION:} It is a criterion for identifying, from a given causal graph, which variables in  the graph must be independent conditional on which other variables. D-separation rule needs to consider three basic causal structures in a DAG \cite{pearl2016causal,pearl2014probabilistic}. These structures correspond to causation, endogenous selection \cite{elwert2014endogenous,alma}, and confounding. We shall use a shorthand notation for conditional independence \cite{dawid1979conditional}. These structures are chains (i.e, $e\rightarrow d\rightarrow f$, the path is d-separated when $e\!\perp\!\!\!\perp f | d$), forks (i.e, $e\leftarrow d\rightarrow f$, the path is d-separated when $e\!\perp\!\!\!\perp f | d$), and inverted forks (i.e, $e\rightarrow d\leftarrow f$ the path is d-connected when $e\not\!\perp\!\!\!\perp f | d$, so to be d-separated we can not condition on $d$ which is a collider). 

We identified the D-separation conditions implied by the causal model and tested the implications to some extent using the dataset. The results are shown in Table \ref{table3}. 

\begin{figure*}[t]
\centerline{\includegraphics[width=15cm,height=6cm]{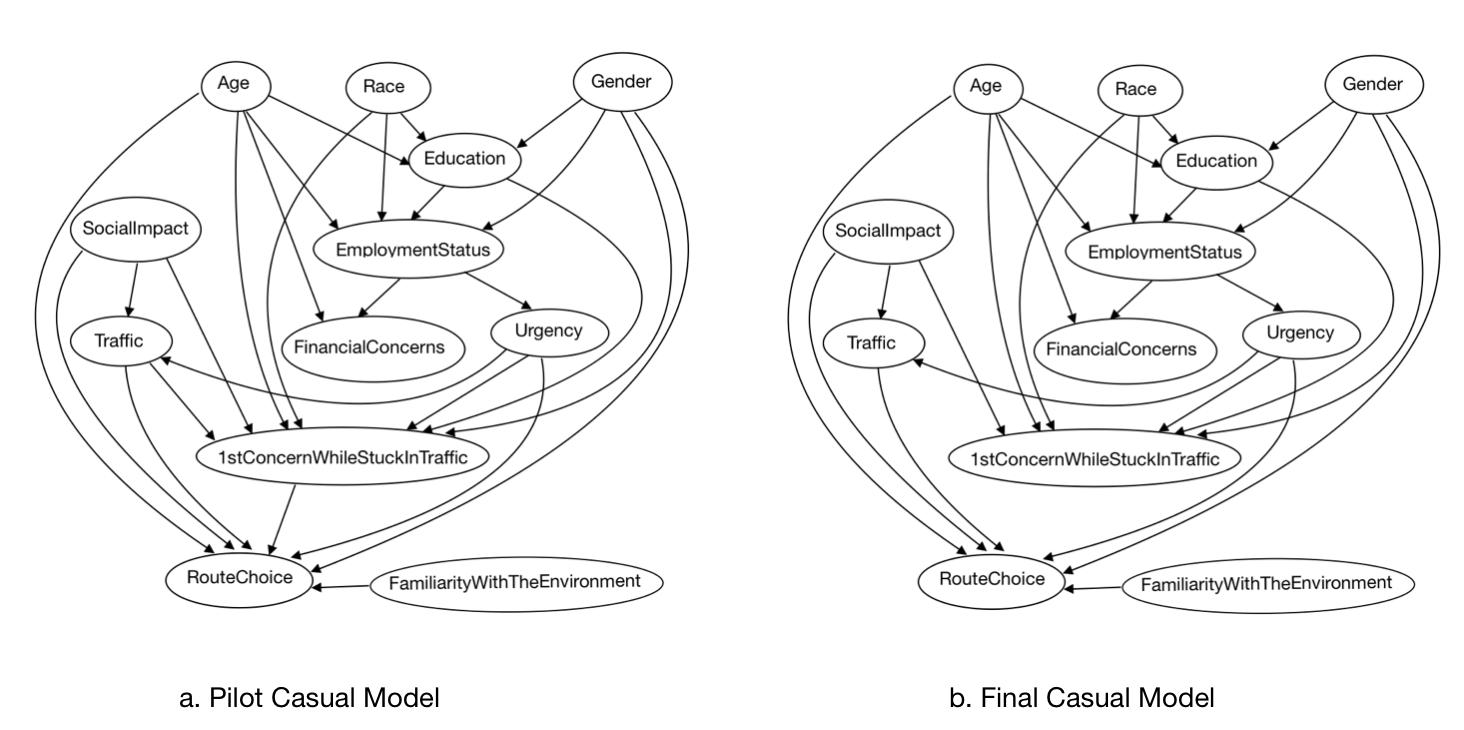}}
  \caption{Casual Models}
\label{fig1}
\end{figure*}

In the conditional independence test, our null hypothesis states that two variables are independent conditional on the other variable. So, if the $p$-value is greater than the significance level $\alpha$ = 0.01, we will accept our null hypothesis; otherwise we will reject it. After the test, it showed that some of the conditional independences implied by causal model were not consistent with the probability distribution underlying the dataset. For example, the result of conditional independence test against  the dataset suggests that the variable of 1st concern while stuck in the traffic and the variable of route choice are independent conditioned on the variable of traffic. However, in the DAG, there is a direct edge between the variable of 1st concern while stuck in the traffic and the variable of route choice which is an indication of dependency. So we will eliminate this edge to make D-separation condition in the model match the conditional independence in the data.

  \begin{table}[h]
\begin{center} 
\caption{Some of the Conditional Independence Test} 
\label{table3} 
\vskip 0.18in
\begin{tabular}{cc} 
\hline
 Conditional Independence  & P-value ($\alpha$=0.01)\\
\hline
1stConcernWhileStuckInTraffic\\ 
and RouteChoice given Traffic& p=0.037 \\
\hline
RouteChoice and 1stConcern-\\
WhileStuckInTraffic given Age & p=0.043\\
\hline
1stConcernWhileStuckInTraffic\\ 
and Education given Race & p=3.57E-10\\
\hline
Education and 1stConcernWhile-\\
StuckInTraffic given Gender &  p$<$2.2E-16\\
\hline
1stConcernWhileStuckInTraffic\\ 
and RouteChoice given SocialImpact & p=0.504\\
\hline
RouteChoice and Traffic given\\
SocialImpact & p$<$2.2E-16\\
\hline
1stConcernWhileStuckInTraffic\\ 
and RouteChoice given Urgency & p=0.481\\
\hline
Traffic and 1stConcernWhile-\\
StuckInTraffic given Urgency & p=1\\
\hline
Education and EmploymentStatus\\
given Gender & p$<$2.2E-16\\
\hline
Education and EmploymentStatus\\
given Age &  p$<$2.2E-16\\
\hline
Education and EmploymentStatus\\
given Race &  p$<$2.2E-16\\
\hline
FinancialConcern and Employment-\\
Status given Age & p$<$2.2E-16\\
\hline

\end{tabular} 
\end{center} 
\end{table}

We concluded that the pilot model is not a good fit for the data set. 
To modify the pilot  model, we could introduce new variables, remove redundant variables, or modify  the relationship between variables by adding or eliminating nodes and edges. Based on the test results, we modified the pilot model by merely eliminating edges from the node of Traffic to the node of 1st concern while stuck in the traffic, and from the node of 1st concern while stuck in the traffic to the node of Route Choice. In the pilot DAG, there were $12$ nodes (variables) and 26 directed edges. In the final causal DAG, the number of nodes remain the same as in the pilot model, but $24$ directed edges  remain. The pilot and final casual models are shown in  Figure \ref{fig1}. 


\section{Causal effect estimation}
\begin{figure}
 \centering{\includegraphics[scale=0.95]{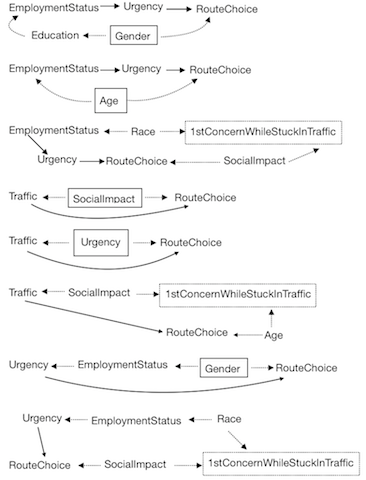}}
\caption{Blocking Backdoor}
\label{fig2}
\end{figure}

 The causal graph shows that between treatment-outcome pairs there is a direct path and an  indirect (back-door) path (i.e, traffic $\rightarrow$ route choice is direct path; traffic $\leftarrow$ social impact $\rightarrow$ route choice is an indirect path).  The back-door path is confounding. When trying to estimate causal effect, we want to block any back-door paths by conditioning on some variables, because such paths are not transmitting causal influences, and if we don't block the back-door path, it confounds the effect that a node has on another node. For instance, as shown in Figure \ref{fig2} (we boxed the collider with  the dashed line and similarly presented the confounder's arrows with dashed line), when trying to calculate the causal effect of employment status on route choice, there exists back-door paths: 1) employment status $\leftarrow$ education $\leftarrow$ gender $\rightarrow$ route choice; the blockage of this path can be ensured by  conditioning on gender which is a confounder; 2) employment status $\leftarrow$ age $\rightarrow$ route choice; the blockage of this path can be ensured by conditioning on the confounder age; 3) employment status $\leftarrow$ race $\rightarrow$ 1st concern while stuck in the traffic $\leftarrow$ social impact $\rightarrow$ route choice; within the path there is a collider which is the variable of 1st concern while stuck in the traffic. So the back-door path is already blocked without conditioning on any variables. However, if we try to condition on the collider we make the path open instead. There are other back-door paths: we haven't listed all of them. After identifying every back-door path between these two variables, we selected age and gender thereby blocking the back-door paths between employment status and route choice. 
 
 To select the variables that entail blockage of the the back-door paths, we carried out graph-surgery as described above. Then, we adjusted these variables to calculate the pure causal effect. We paired every treatment with the outcome variable, identified back-door paths between them, and selected the confounding variables using the Back-door criterion. Between the variables of Urgency, Gender, Race, Age, SocialImpact, FamiliarityWiththeEnvironment, and the RouteChoice there doesn't exist any back-door path. In addition to that, from the variables of 1stConcernWhileStuckInTheTraffic, and FinancialConcern, there  doesn't exist any direct causal path to the variable of RouteChoice. So, there is no casual effect of these two variables on the variable of RouteChoice. The relationship between them can be interpreted as association instead of causation. Hence,  there is no need to estimate the  causal effect of these two variables. We listed the confounders in the the back-door paths between the variables of Traffic, Urgency, Education, EmploymentStatus and RouteChoice (Shown in Table \ref {table4}). Normally,  the strategy of putting in all possible confounders is usually used. However, this strategy may end up adjusting for colliders and mediators that can introduce bias. For example, employment status $\rightarrow$ urgency $\rightarrow$ route choice. In this direct path, urgency is mediator.  if we condition on the mediator, we will bias our estimate. 
 
  \begin{table}[h]
\begin{center} 
\caption{Confounder in Backdoor Paths} 
\label{table4} 
\vskip 0.12in
\begin{tabular}{cc} 
\hline
 Variables  & Confounder\\
\hline
Traffic & Urgency, SocialImpact   \\
\hline
Urgency     & Gender, Age \\
\hline
Education          & Race, Gender, Age \\
\hline
EmploymentStatus & Gender, Age     \\
\hline
\end{tabular} 
\end{center} 
\end{table}
   
 We applied Inverse Probability(IP) weighting method to adjust the variables $Z$, which are the confounders. The purpose of using IP weighting is to break the association between the covariates $Z$ and treatment $X$ to estimate true causal effect on outcome variable $y$ \cite{hernan2006estimating,robins1994estimation}. By predicting $Z$ based on $X$, we can estimate the propensity score $\Pr (X=x | Z)$. We can get a propensity score using non-parametric (i.e, probability) or parametric methods (i.e, regression model). If the data is high-dimensional with many covariates and some of them with multiple levels, it is desirable to use a parametric method. In our study, we have 12 variables and some variables have more than two levels. To find propensity score, we applied  logistic regression model. The equation is given below: 
  \begin{equation} \displaystyle \Pr (X_i=x | Z_i) =\dfrac{1}{1+ e^{-\beta Z_i} }\end{equation} 
 
 After getting propensity scores, we used them to obtain the weights $W$ to create a pseudo-sample in which there is no association between the covariates and treatment. The IP weighting formula is given below:
 \begin{equation} W_i =\dfrac{X_i}{\Pr (X|Z_i)} + \dfrac{1-X_i}{1- \Pr (X|Z_i)} \end{equation}
 where $X_i$ indicates if the $i$th subject was treated.

We started our approach of calculating the causal effect by training a model with covariates $Z$ to predict $X$. Our treatments $X$ are categorical variables, so we calculated the propensity scores $\Pr (X|Z)$ by applying equation 2.

After estimating the propensity scores, we applied equation 3 to obtain the $IP$ weight. We used stabilizing factor $\Pr(X)$ in the numerator to narrow the range of the $\Pr(X)/W$. After we obtained stabilized IP weights as $SW=\Pr(x)/W$, we trained new model with treatment variables $X$ as features and outcome Y by using $SW_i$ as sample weight for the $i$th observation. Then, we used this model to predict the causal effect. In this study, outcome variable is categorical data, so we used logistic regression again to obtain the casual odds ratio as a casual effect measure.

  \begin{table}[h]
\caption{Average Casual Estimation Result} 
\label{table5} 
\vskip 0.15in
\begin{tabular}{ccc} 
\hline
Variables & ATE  &95\%Conf.Interval \\
\hline
SocialImpact& \\
  yes vs no &4.945&3.966 --  5.919\\
  \hline
  Urgency&\\
  Urgent vs Unurgent& 1.193& 0.773 -- 1.612\\
  \hline
  Age&\\
  Middle vs Young &1.081& 0.179 -- 1.982\\
  Old vs Young & 0.461& -3.274 -- 4.196\\
  \hline
  Gender&\\
  Male vs Female & 0.976& 0.135--1.816\\
  \hline
  Race&\\
  Middle Eastern vs White & 5.760& 4.587 -- 6.932\\
  Other vs White & 3.966& 2.750 -- 5.181\\
  \hline
  EmploymentStatus &\\
  PartTime vs Unemployed & 0.007& -6.049 -- 7.651\\
  FullTime vs Unemployed & 0.015& -4.140 -- 4.213\\
  Student vs Unemployed & 0.010& -4.471 -- 4.582\\
  \hline
  Education &\\
  HighSchool vs PostGraduate &0.054& -2.431 -- 2.539\\
  College vs PostGraduate & 1.231& 0.825 -- 1.636\\
  \hline
  Traffic&\\
  Medium vs Normal & 3.663& 1.953 -- 5.372\\
  Heavy vs Normal & 6.562& 4.817 -- 8.306\\
  \hline
  FamilarityWithEnvironment&\\
  OnceAMonth vs OnceAWeek &2.795& 0.938 -- 4.651\\
  OnceAYear vs OnceAWeek & 2.778& 1.374 -- 4.181\\
  \hline
  
\end{tabular} 
\end{table}
 \begin{figure}
\includegraphics[scale=0.4]{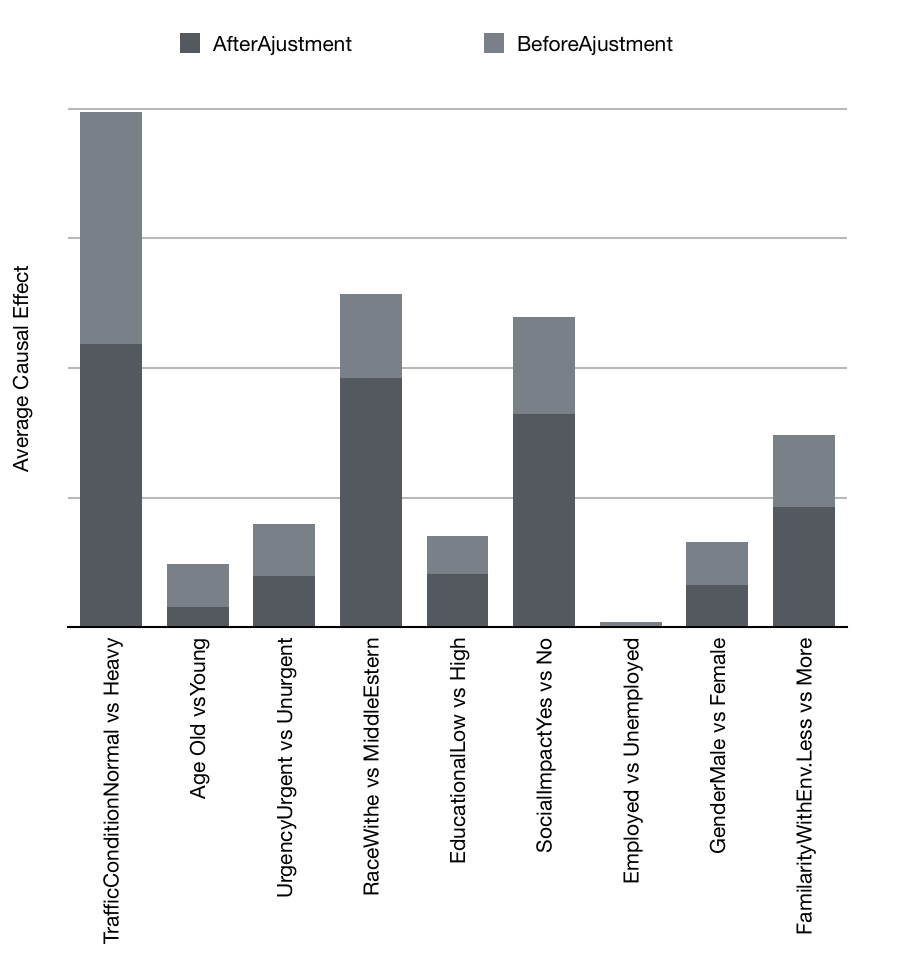}
\caption{Average Causal Effects Of Variables Computed by Adjusted and Un-Adjusted Regression Model}
\label{figure3}
\end{figure}
 Based on our determination of average causal effect (shown in Table \ref{table5}), the result suggests that when heavy and medium traffic conditions are compared with the normal condition separately, their effects have significant magnitude. It implies that  changes in traffic conditions impact drivers route choice. More specifically, when the traffic is heavy, proportion of drivers who choose the nearest exit is about $6$ times greater than that when traffic is normal. However, when the traffic is medium, proportion of drivers who choose the nearest exit is about $3$ times greater than that  when traffic is normal.  So, we can conclude that when the traffic is normal, the drivers are more likely to stay on the high way. Considering  social impact, a driver would be influenced by watching other drivers take the exit. The proportion of drivers who choose the nearest exit is about $5$ times greater when they get influenced than they don't. Considering familiarity with environment, proportion of drivers, who are not familiar with the road, choosing the nearest exit is about $3$ times greater than those who are familiar with the road. Considering race, white people are more likely to stay on the highway than middle eastern people or others.  Age and Urgency also have significant effect on drivers route choice.  
 We also conducted another experiment in which we built an estimator for route choice without adjusting  for confounding factors, and compared the results with the one of causal inference (shown in Figure \ref{figure3}).  The results suggest that on the un-adjusted estimator, the effect of age and  employment status are overestimated, race, social impact,  and familiarity with the environment are under estimated. This is because there are confounding and collider sources between the path of these variables and the outcome. Based on causal inference, the effect of traffic, race, social impact, and familiarity with the environment are more significant than others.

 \section{Conclusions}

 This paper described a causal analysis of route choice behavior of drivers using data collected from a Stated Choice Experiment in an Immersive Virtual Environment (IVE).  This work  will not only fill in the lack of causality based approaches  in the transportation field, but it also showed that without adjustment on treatment, causal effect results will be affected by spurious correlation as well.
 
\section*{Acknowledgment}
This research was supported by Transportation Consortium of South-Central States (Tran-SET) Award No 18ITSLSU09/69A3551747016. Any opinions, findings, and conclusions or recommendations expressed in this material are those of the author(s) and do not necessarily reflect the views of the sponsor. 
 
\bibliographystyle{apacite}

\setlength{\bibleftmargin}{.125in}
\setlength{\bibindent}{-\bibleftmargin}


\end{document}